\documentclass{article}
\usepackage{arxiv}

\usepackage[utf8]{inputenc} 
\usepackage[T1]{fontenc}    
\usepackage{hyperref}       
\usepackage{url}            
\usepackage{booktabs}       
\usepackage{amsfonts}       
\usepackage{nicefrac}       
\usepackage{microtype}      
\usepackage{lipsum}
\usepackage{eurosym}
\usepackage{graphicx}
\usepackage{cite}
\graphicspath{ {./images/} }
\newcommand{\ra}[1]{\renewcommand{\arraystretch}{#1}}
\usepackage{listings}
\usepackage{color}
\usepackage{alltt}
\usepackage{booktabs,caption}
\usepackage[flushleft]{threeparttable}

\definecolor{dkgreen}{rgb}{0,0,0}
\definecolor{gray}{rgb}{0,0,0}
\definecolor{mauve}{rgb}{0,0,0}
\title{A step-by-step guide to design, implement, and analyze a discrete choice experiment}

\author{
 Daniel Pérez-Troncoso \\
  Faculty of Economics\\
  University of Granada\\
  \texttt{danielperez@ugr.es} \\
}

\begin{document}
\maketitle
\begin{abstract}
Discrete Choice Experiments (DCE) have been widely used in health economics, environmental valuation, and other disciplines. However, there is a lack of resources disclosing the whole procedure of carrying out a DCE. This document aims to assist anyone wishing to use the power of DCEs to understand people's behavior by providing a comprehensive guide to the procedure. This guide contains all the code needed to design, implement, and analyze a DCE using only free software.
\end{abstract}


\section{Introduction}
Discrete choice experiments (DCE) have been widely used in health economics\cite{1,2,3,4}, environmental valuation\cite{5,6}, and other disciplines where there is not a market institution where consumers’ decision can be observed. For addressing the lack of information, in a DCE respondents are presented with different alternatives to choose from (simulating a market situation). Each alternative comprises several attributes and levels to force consumers to make trade-offs according to their preferences\cite{7,8}. Despite the usefulness of discrete choice experiments, they can be hard to design, implement, and analyze, and usually paid software is recommended to carry out the research. This document is aimed to elaborate on the full process of studying consumers’ preferences through a discrete choice experiment.

\section{Elements in a discrete choice experiment}
\label{sec:headings}
Before starting with the steps of the procedure, it is worth to understand the parts and nomenclature used to design the elements of a DCE. 

A discrete choice experiment is usually administered through a questionnaire than contains [at least] three parts: a) an introduction, 2) the DCE itself, 3) respondent information \cite{1}. The introduction is necessary to make respondents understand what they are responding to, why are they doing that, and how to do it correctly. The second section, the most important one, contains the DCE itself. In this section, respondents repeat a choice task (Example in Table~\ref{tab1})  between different hypothetical scenarios as many times as the analyst decides. These choice tasks are known as 'choice sets', and they usually contain 2 or 3 alternatives. Attributes and levels are the main components of the choice sets. Attributes are the categories where the characteristics of the hypothetical product are classified (see the left column in Table~\ref{tab1}). Levels are the different functionalities of the product classified by attributes. Following the example in Table~\ref{tab1}, a possible selection of attributes and levels is presented in Table~\ref{tab2}.

\begin{table}
 \caption{Example Choice Set}
  \centering
  \begin{tabular}{llll}
    \toprule
    Attributes     & Option 1     & Option 2      & Option 3  \\
    \midrule
    Efficacy       & 100\%        & 80\%          &           \\
    Side effects   & nausea in 1 out of 10,000 patients. & none &\\
    Dose           & Once a week & Once a day     &           \\
    Administration & Injection   & Oral           &         \\
    Price          & 150\euro/month  & 200\euro/month    & 0\euro \\
    \toprule
     & Choose option 1 	 & Choose option 2 	 & Choose option 3  \\
     \bottomrule
  \end{tabular}
  \label{tab1}
\end{table}

In the third section, respondents can be asked about anything relevant to the analysis of their decisions. Usually, this section includes socio-demographic questions (like income, age, education, and gender).

\begin{table}
 \caption{Example selection of attributes and levels}
  \centering
  \begin{tabular}{lll}
    \toprule
    Attributes     & Levels     & No. of levels      \\
    \midrule
    Efficacy       & 1) 80\%, 2) 90\%, 3) 100\% & 3     \\
    Side effects & 1) nausea in 1 out of 10,000 patients, 2) none & 2 \\
    Dose           & 1) Once a day, 2) Once a week, 3) Once a month & 3\\
    Administration & 1) Oral, 2) Injection   & 2                    \\
    Price          & 1) 100\euro/month, 2) 150\euro/month, 3) 200\euro/month  &  3   \\
     \bottomrule
  \end{tabular}
  \label{tab2}
\end{table}

If the attributes and levels in Table~\ref{tab2} were combined to create all the possible choice sets, they would lead to $3^2\times2^2=108  $ different profiles. Pairing the profiles we would have a total of $(108\times107)/2=5778$ different choice sets. A DCE using all the existing combinations is known as a full factorial design. However, most cases we do not have such a large sample, so "the central question is then how to combine the alternatives from the full factorial design into the choice sets so that a maximum amount of information is extracted [...]" \cite[p.~284]{9}. In this document, we will address an optimization technique to obtain the maximum possible information from our design. 
\section{Attributes and levels selection}

Attributes and levels selection is an essential part of the DCE design. Poor selection will lead to invalid results, so it is worth doing extensive qualitative research before making the final selection. Some useful references are the WHO guide\cite{2}, the Gerard et al. manual\cite{4}, and the ISPOR checklist\cite{10}. A literature review will be essential but can be supplemented by consultation with experts and stakeholders. A common strategy here is to gather all possible attributes and levels into one long list and review it by discarding some and merging others. As a general rule, the selection of attributes and levels should be justifiable by evidence.
As the number of attributes and levels increases, each response provides less information, so a balance needs to be found between specification and efficiency. Besides, a limited selection of attributes and levels is recommended so as not to confuse the respondent. 70\% of existing DCEs use from 3 to 7 attributes\cite{1}.

Conversely, an incomplete specification of attributes and levels could lead to a large estimation error (note that the results of this methodology will be analyzed in the random utility framework \cite{11}). According to this theory, the utility that an individual $n$ gets from a product $j$ is given by a deterministic component, $V_{nj}$, and a random component not observed by the researcher, $\epsilon_{nj}$ .

\begin{equation}
u_{nj}=V_{nj}+\epsilon_{nj}
\end{equation}

In Equation 1, the deterministic component, $V_{nj}$, is composed of the attributes' levels and its coefficients, $\beta_{nj}x_{nj}$. As higher the number of attributes and levels is, the lower the error component, $\epsilon_{nj}$ and vice versa. That is why we must balance simplicity and precision. 

Finally, when deciding the attributes and levels in a DCE we have to include a price attribute with two or more cost or price levels. This attribute is vital, as respondents will make most decisions with it in mind. It will allow us to measure the trade-offs between functionalities (levels) and monetary units. Thus, we can measure willingness to pay (WTP) as a simple division of the level coefficient by the price attribute coefficient. 

\section{Pilot study and experimental design}
The efficient design of a DCE requires prior information in the form of "prior coefficients" of the model we want to estimate to perform an optimization procedure. This procedure consists of creating a design matrix (experimental design) that reduces the variance of the coefficients that we are going to calculate\cite{9}. One of the most common ways of estimating this variance is
\begin{equation}
d-error=|\Omega^{1/k}|
\end{equation}
where $\Omega$ is the covariance matrix of the model coefficients, and $K$ is the number of parameters of the model. Thus, an algorithm can be employed to iterate random designs matrices until it finds the experimental design with lower D-error. This procedure ensures the efficiency of the design and the accuracy of the estimators. 

A pilot study consists of creating an efficient design based on a set of prior coefficients equal to zero and presenting the resulting design to a limited number of respondents to achieve the first set of actual prior coefficients. Once we achieved the coefficients, we can use them to generate another efficient design.

The first step is to design a pilot DCE with priors equal to zero. To do it, we will use R (https://www.r-project.org/), (optional but recommended) RStudio (https://rstudio.com/), and the package 'idefix' \cite{17}. 

\begin{verbatim}
    1. install.packages(“idefix”)
    2. library(idefix)
    3. levels <- c(3,2,3,2,3)
    4. coding <-c("E","E","E","E","E")
\end{verbatim}
The first and second line will install and activate the package that we are going to use to create an efficient design. At line 3, we are creating a vector where each element is an attribute, and its value is the number of levels in that attribute. At line 4, we specify another vector with the type of coding that we are going to use in each attribute. In this case, we will use "effects coding" so we write an "E" per attribute. If we want to see the different alternatives from all attributes and levels combination, we can create and display the profiles:
\begin{verbatim}
    5. Profiles (lvls=levels, coding=coding)
\end{verbatim}
The input at line 5 will generate all different alternatives from our design. In this case, we can see that 108 profiles were generated: $3^3*2^2=108$.

Next, we want to generate the D-efficient design. To do so, we will use Fedorov modified algorithm\cite{13} included in the -idefix- package. This algorithm generates a random initial design from the set of profiles and randomly switches the levels and compares the D-error\cite{14}. This process goes on until the iteration $n-1$ and $n$ exhibit the same D-error. By reducing D-error we are getting close to the principles of good DCE design: orthogonality, level balance, minimal overlap, and utility balance \cite{15}.

As we mentioned before, because this is the first pretest, we do not have any prior information. Thus, when specifying the vector with the prior coefficients, we need to set them all to zero. 
\begin{verbatim}
    6. priors <- c(0, 0, 0, 0, 0, 0, 0, 0, 0)
    7. s <- diag(length(priors))
    8. sim <- MASS::mvrnorm(n = 500, mu = priors, Sigma = s)
    9. sim <- list(sim[, 1:1], sim[, 2:9])
    10. alt.cte <- c(0, 0, 1)
    11. d<-CEA(lvls=levels, coding=coding, n.alts=3, n.sets=16, alt.cte=alt.cte, 
        par.draws=sim, no.choice=TRUE, best=TRUE)
\end{verbatim}
At line 6, we have a total of 9 zeros, although we have 13 coefficients. That is because we are omitting a level per attribute (due to the effects coding) and adding an additional zero as the coefficient of the dummy representing the opt-out alternative. Thus, the amount of coefficients in the priors vector is $(l-k)+1$ where $l$ is the number of levels and $k$ is the number of attributes.

Lines 7 and 8 are part of a simulation procedure where 500 random draws are obtained from a normal distribution with mean equal to the priors specified. Those results will be used to simulate a response pattern and generate a design that reduces the estimators' variance. At line 9, we create two lists, the first one for the coefficient of the dummy variable in the opt-out alternative, and the other for the rest of the coefficients. At line 10, we create a vector indicating the position of the opt-out alternative and, finally, line 11 produces an output in $d$ with the D-efficient design. As we can see, we are creating a D-efficient design based on the priors stored in $priors=(0,0,0,0,0,0,0,0,0)$, with five attributes with, respectively, $3,2,3,2,3$ levels. We are using effects coding, three alternatives per choice set (the third one is a constant no-choice/opt-out), and sixteen choice sets per respondent. 
\begin{verbatim}
    12. design <- d$design
\end{verbatim}
At line 12, we store the design matrix in the variable 'design'. Once we have the design matrix, we need to interpret it to obtain the choice sets. Our design matrix is an $n \times m$ matrix where $n$ is the number of choice sets times the number of alternatives, and $m$ is equal to the number of coefficients in the priors vector, $(l-k)+1$. In this design, we are using 'effects coding' where all levels are written as dummy variables except for one omitted level. 

Now we want to interpret the design matrix to understand how alternatives are distributed in across the choice sets. In Table~\ref{tab3} we can see how, depending on the coding used, the variables take one value or another to represent one attribute's level.  
\begin{table}\centering
\caption{Dummy vs. effects coding}
\ra{1.3}
\begin{tabular}{@{}rrrrcrrr@{}}\toprule
& \multicolumn{3}{c}{Dummy coding} & \phantom{abc}& \multicolumn{3}{c}{Effects coding} \\
\cmidrule{2-4} \cmidrule{6-8}
& $\beta_2$ & $\beta_3$ & $\beta_4$ && $\beta_1$ & $\beta_2$ & $\beta_3$\\ \midrule\\
Level 1 & 0 & 0 & 0 && 1 & 0 & 0 \\
Level 2 & 1 & 0 & 0 && 0 & 1 & 0 \\
Level 3 & 0 & 1 & 0 && 0 & 0 & 1 \\
Level 4 & 0 & 0 & 1 && -1 & -1 & -1 \\
\bottomrule
\end{tabular}
\label{tab3}
\end{table}

For example, figure one shows the output of $design$. In this case, the choice set one has alternative 1 with levels $(1,1,3,2,3)$ vs. alternative 2 with levels $(3,2,1,1,3)$ versus the "none of them option". The second choice set contains alternative 1, $(1,1,1,2,3)$, and alternative 2, $(2,2,2,1,1)$.

\begin{figure} 
\caption{Screenshot to design matrix in R}
    \centering
    \includegraphics{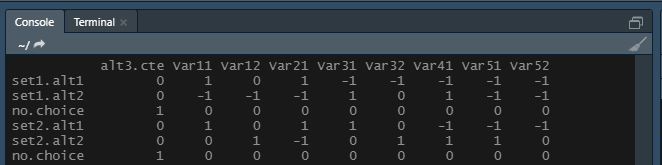}
\end{figure}

\section{Survey design}
DCEs are commonly administered via online surveys. In this section, we will focus on how to design an online survey. Our survey will be divided into three parts: 1) introduction, 2) DCE, and 3) respondent information. For this example, in the introduction, we will add an introductory text, explaining to the respondents how the experiment works, and a timer that will not let them skip the introduction until after a minute. To do this, we will create two HTML documents with the following code: 

\begin{lstlisting}
<!DOCTYPE HTML>
<html>
  <head>
    <meta charset="utf-8">
    <meta name="viewport" content="width=device-width, initial-scale=1, shrink-to-fit=no">

    <title>Title in the tab</title>

    <script src="http://code.jquery.com/jquery-latest.min.js" type="text/javascript"></script>
    <link rel="stylesheet" href="https://stackpath.bootstrapcdn.com/bootstrap/4.5.0/css/bootstrap.min.css" integrity="sha384-9aIt2nRpC12Uk9gS9baDl411NQApFmC26EwAOH8WgZl5MYYxFfc+NcPb1dKGj7Sk" crossorigin="anonymous">
  </head>
  
  <body>
    <h1>Welcome to my survey</h1>
    <p>Lorem ipsum dolor sit amet, consectetur adipiscing elit, sed do eiusmod tempor incididunt ut labore et dolore magna aliqua. Ut enim ad minim veniam, quis nostrud exercitation ullamco laboris nisi ut aliquip ex ea commodo consequat. Duis aute irure dolor in reprehenderit in voluptate velit esse cillum dolore eu fugiat nulla pariatur. Excepteur sint occaecat cupidatat non proident, sunt in culpa qui officia deserunt mollit anim id est laborum.</p>
   
    <p>Before you go to the survey, you need to wait 1 minute so please use this time to read the introductory text<p></p>
    <button id="exit" onclick="document.location='redirect.html'">Go to survey!</button>
    
    <script>
      $(document).ready(function() {
        $("#exit").prop("disabled", true);
        setTimeout(function(){$("#exit").prop("disabled", false)}, 60000);
      });
    </script>
  </body>
</html>

\end{lstlisting}

The code above can be copied and pasted into a plain text editor and saved as .html. The dummy text in $<p></p>$ can be modified according to the necessities of the study, same with the $60000$ value (equivalent to 1 minute) in the $setTimeout$ function.

The link on the "Go to Survey" button redirects to another .html file whose function is to send the respondent randomly to an A or B link. Thus, the code below will be useful when the survey is blocked into two parts and we want to send respondents randomly to block A or B. 

\begin{lstlisting}
<html>
<head>
    <title></title>
    <meta http-equiv="Content-Type" content="text/html; charset=iso-8859-1">
    <script type="text/javascript">
      var pageArr = ["https://cran.r-project.org/", "http://gretl.sourceforge.net/"];
        document.location.href = pageArr[Math.ceil(Math.random()*2)-1];
    </script>
</head>

<body>

</body>
</html>

\end{lstlisting}
This code needs to be saved and named redirect.html. In the example code, the page randomly redirects to either the R page or the Gretl page. Those links can be modified with the links of both parts of the survey to create the random assignment. The introduction and the redirect pages need to be uploaded to the Internet as a new webpage or part of an existing domain.

Next, we will design the blocks in Google Forms. The optimal option here is Qualtrics, but this is a paid software which only some universities and companies have access too. Although Google Forms has many limitations, we can expect it to remain always free. Thus, we will create two Google Forms, one for each block. In each form we need to:
\begin{itemize}
    \item click on Settings / Presentation / Shuffle question order,
    \item create two sections,
    \item create the socio-demographic questions in section 2, 
    \item create the choice sets in section 1.
\end{itemize}
To create the choice sets best option might be using 'multiple choice' questions and adding an image to each question. That image can be designed at any text editor and could be something similar to Table ~\ref{tab2} (icons can be added to improve respondent's interpretation).

\section{Analysis of results}
The results of our discrete choice experiment will be analysed in line with McFadden’s Random Utility Theory (RUT) \cite{11}. According to RUT, the utility that an individual $n$ gets from an alternative $j$, $u_{nj}$, is given by a deterministic component, $V_{nj}$, plus an unobserved component, $\epsilon_{nj}$ \cite{19}: 

\begin{equation}
    u_{nj}=V_{nj}+\epsilon_{nj}
\end{equation}

Additionally, the deterministic component of the utility, $V_{nj}$, can be specified as the observed characteristics of the alternatives (levels), $x_{nj}$, related to the value that respondents assign to each level, $\beta_{nj}$. We can specify this relationship as
\begin{equation}
    u_{nj}=\beta_{nj} x_{nj}+\epsilon_{nj}, 
\end{equation}

where the value of the betas can be estimated according to different assumptions. For instance, if we assume that there is not taste heterogeneity among respondents, the equation can be specified as

\begin{equation}
    u_{j}=\beta x_{j} + \epsilon_{j}
\end{equation}

where the utility does not vary among respondents but does vary among alternatives (note the deletion of the sub-index n). This model is known as the conditional logit model (CLM) \cite{18}, and it results in the same coefficients for all respondents. While the CLM is the easy and reliable model recommended by the ISPOR task force report \cite{18}, it has strong assumptions (IIA) and some limitations (it assumes preference homogeneity and ignores the panel nature of the data i.e. do not consider the 'personid' variable.) 

To avoid the limitations of the CLM, we might want to take into account taste heterogeneity among consumers. In this case, the utility will be specified as 

\begin{equation}
    u_{nj}=\beta_{nj}x_{j}+\epsilon_{nj}
\end{equation}

where we compute a coefficient per respondent (note that only $x_j$ lacks the $n$ sub-index because all respondents are presented with the same levels). This specification can be estimated with the mixed logit model (XLM). The XLM is a flexible model adaptable to any choice situation \cite{19}. In this model, the coefficients are computed per respondent over a simulated density f($\beta$). Thus we are computing something similar to a CLM per respondent. When the model converges, it reports the mean of the $\beta$s of each level along with its standard deviation. Thus, we can know whether or not differences among respondents parameters are significant. 

\subsection{Reshaping the data}
We need to resize the data from the Google Forms survey. When we download the survey data from Google Forms, the data is shaped in ‘wide’ format (Table ~\ref{widetab}), and we need to resize it to ‘long’ format (Table ~\ref{longtab}) for the logit model to understand it. Since we have blocked the survey into two parts, we will do this individually for each block. To do this, we will use the following commands.

\begin{table}
\centering
   \begin{threeparttable}
 \caption{Wide shaped data set}
  \begin{tabular}{llllll}
    \toprule
    
     & 1   & 2   & 3   & 4   & 5  \\
    
    \midrule
    1          & A   & B   &  A  &  C  & B      \\
    2     & B    & C   &  A  &  B   & B     \\
    3     & C   & C    & A   & A & C    \\
    4   & A & A & C & B & A\\
    ...  & ... & ... & ... & ... & ...\\ 
    N  & A   & B  &  B & C & C    \\
     \bottomrule
  \end{tabular}
   \begin{tablenotes}
      \small
      \item Choice sets in columns, responses in rows.
    \end{tablenotes}
    \label{widetab}
  \end{threeparttable}
\quad 
\quad
\quad
\quad
   \begin{threeparttable}
 \caption{Long shaped data set}
  \begin{tabular}{llllll}
    \toprule
    
    personid & cs   & choice   & alt   & ...   & price  \\
    
    \midrule
    1     & 1 & 1 & 1  &  ...  & 150 \\
    1     & 1 & 0 & 2  &  ...  & 100  \\
    1     & 1 & 0 & 3  &  ...  & 200  \\
    1     & 2 & 0 & 1  &  ...  & 200\\
    1     & 2 & 1 & 2  &  ...  & 100\\
    1    & 2 & 0 & 3  &  ...  & 150\\
    ...  & ... & ... & ... & ... & ...\\ 
    50  &  800  & 0 &  3 & ... & 100    \\
     \bottomrule
  \end{tabular}
   \begin{tablenotes}
      \small
      \item Respondent id, choice set id, choice dummy, alternative id, and alternative specific variables (levels) in columns. For example, respondent 1 chooses alternative 1 (option A) in choice set 1.
    \end{tablenotes}
    \label{longtab}
  \end{threeparttable}
  \end{table}

\begin{verbatim}
    12. data<-data[2:ncol(data)]
    13. personid <- rownames(data)
    14. personid <- as.integer(personid)
    15. data <- cbind(personid, data)
\end{verbatim}

First, we remove the “timestamp” variable (12), then we need to create a person id (13, 14, 15). At this point, we have our dataset as in Table 4 where columns represent the different choice sets. It is convenient having the same id for the choice sets than in the design matrix (for example, column 1 in Table 4 is the response of respondent 1 to set 1 in Figure 1). Then, the library -reshape2- will help us to switch from ‘wide’ to ‘long’ shaped format (16, 17, 18). 

\begin{verbatim}
    16. install.packages(“reshape2”)
    17. library(reshape2)
    18. data <- melt(data, id.vars=c(“personid”))
\end{verbatim}

Once we have our dataset reshaped, we need a row per alternative. To do this, we will multiply the number of choice sets (which, at this point, coincides with the number of rows per respondent) per the number of alternatives. In our case, we have three alternatives per choice set (19, 20). We also need an alternative id (21, 22, 23) and a choice set unique id (24, 25, 26).

\begin{verbatim}
    19. data <-rbind(data, data, data)
    20. data <- data[order(data$personid, data$variable),]
    21. x <- nrow(data)/3
    22. alt <- rep(1:3, x)
    23. data <- cbind(data, alt)
    24. cs <- rep(1:x, each= 3)
    25. cs <- sort(cs)
    26. data <- cbind(data,cs)
\end{verbatim}

Next, the package -dplyr- is needed for the final steps (27, 28). At line 29, we are creating the choice variable that will take value 1 if the alternative in its row is selected or 0 otherwise. This command only works if the response variable was coded as “A”, “B”, “C” for options A, B, C. If not, it has to be modified for it to work. 

\begin{verbatim}
    27. install.packages(“dplyr”)
    28. library(dplyr)
    29. data <- mutate(data, choice=ifelse(value == "A" & alt=="1" | value== "B" & alt=="2"
        | value== "C" & alt=="3", 1, 0))
\end{verbatim}

Finally, we have to add the alternative specific variables directly from our design matrix. First, we need to select the part of the design matrix that correspond to our block. For instance, if we are working with Block 1 with choice sets 1-8, we want to split the design matrix into two parts (cs 1-8, 9-16) \footnote{The choice sets id in the database must correspond with the choice sets id in the design matrix. For example, responses in column '1' of Table~\ref{widetab} need to be responses to choice set '1' of the design matrix. Once data has been reshaped from wide to long format, we can bind the response data (with 8x3=24 rows per respondent) to the design matrix (which divided into two groups has 24 rows). Thus, responses to choice sets 1-8 will be binded to the first 24 rows of the design matrix while responses to choice sets 9-16 will be binded to the last 24 rows of the design matrix.}.

\begin{verbatim}
   30. design1 <- [1:24,]
   31. design2 <- [25:48,]
\end{verbatim}

Thus, ‘design1’ contains the first 8 choice sets while ‘design2’ contains from the choice set 9 to the choice set 16. To add the alternative specific variables to each response first, we adapt the ‘design1(2)’ to the number of responses (line 32), and then we merge responses and design (line 33).

\begin{verbatim}
    32. design1 <- design1[rep(seq_len(nrow(design1)), length(personid)), ]
    33. final1<- cbind(data, design1)
\end{verbatim}

Once this process has been carried out in both parts of the design (final1 and final2), we can merge both datasets: 

\begin{verbatim}
    34. final <- rbind(final1, final2)
\end{verbatim}
\subsection{Computing and reporting the CLM}
Now we can estimate the conditional logit model. Typing lines 33, 34, and 35, we will have enough information to report the results as in Table~\ref{tab6}. As we can see from the syntax at line 34, the model output does not contain the coefficient of omitted levels (100\% efficacy, no side effects, etc.). However, we can compute it as the negative sum of the rest of the coefficients in the same attribute.  
\begin{verbatim}
    33. library(survival)
    34. resultsCLM <- clogit(choice~alt3.cte + Var11 + Var12 + Var21 + Var31 + Var32 +
        Var41 + Var51 + Var52 + strata (cs), data(final)
    35. summary(resultsCLM)
\end{verbatim}

In the CLM, we cannot include respondents characteristics since they do not vary among alternatives. The coefficients that we obtain are associated with the effect (or 'weight') of the level in the decision making process in relation to the mean attribute effect \cite{18}.  We can improve the interpretation of the conditional logit model by plotting the coefficients of the regression as done in Figure 2. For example, in this case respondents obtain less utility from increasing efficacy from 90\% to 80\% (+0.85) than from 90\% to 100\% (+3.13). However, the most important attribute is 'Dose' because the difference between its lowest and highest level is greater than the difference in 'Efficacy' (5.27 vs. 3.98).

\begin{figure}
    \centering
    \includegraphics[scale=0.16]{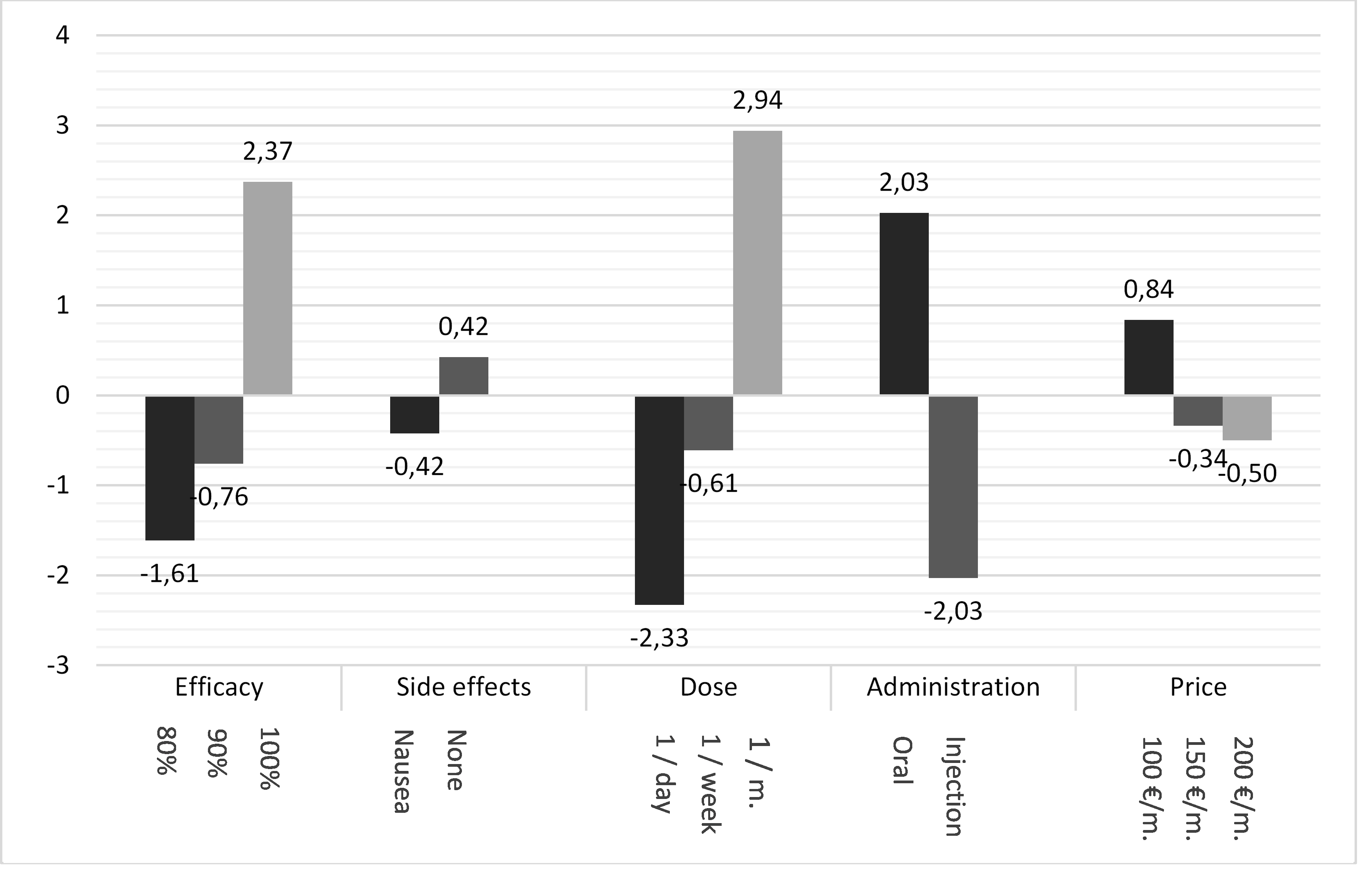}
    \caption{CLM results}
    \label{fig:clm}
\end{figure}

\subsection{Computing and reporting the XLM}

To estimate the XLM, we need to set up out data set by typing the following code: 
\begin{verbatim}
    36. D <- dfidx(final, choice="choice", idx = list(c("cs", "personid"), "alt"), idnames 
        = c("cs", "alt")
    37. resultsXLM <- mlogit(choice ~ alt3.cte + Var11 + Var12 + Var21 + Var31 + Var32 +
        Var41 + Var51 + Var52 | 0, data=D, rpar=c(Var11="n", Var12="n", Var21="n", 
        Var31="n", Var32="n", Var41="n", Var51="n" Var52="n"), R=100, halton=NA, panel=TRUE)
\end{verbatim}

At line 36 we are shaping the data for the -mlogit- function to read it properly (using the same variable names that we used along this document should work). At line 37 we are computing the XLM; the -rpar- option allow us to introduce a vector with the coefficients that we want to be random and its distribution ("n" for normal). 

Unlike the CLM, the XLM reports the standard deviation of the coefficients. If the standard deviation is significant, there is taste heterogeneity among respondents. 

\begin{table}
 \caption{Results of the CLM}
  \centering
  \begin{tabular}{llllll}
    \toprule
    Attributes  & Levels   & Coefficient   & SE      & p-value  \\
    \midrule
    Efficacy     & 80\%    & -1.61188   &  0.11955    & p<0.0000      \\
                 & 90\%    & -0.75922   &  0.08631     & p<0.0000      \\
                 & 100\%   & 2.3711     &  (Omitted level)      \\
    Side effects   & Nausea 1/10,000 patients & -0.42223 & 0.06407  & p<0.0000\\
                   & None & 0.42223 & (Omitted level)\\ 
    Dose  & Once a day   & -2.32870   &  0.14545  & p<0.0000    \\
          & Once a week  & -0.61081 & 0.08389  & p<0.0000   \\
          & Once a month  &  2.93951 & (Omitted level) \\
    Administration & Oral & 2.02899 & 0.11731  & p<0.0000  \\
                   & Injection  & -2.02899 & (Omitted level)         \\
    Price      & 100\euro/month  & 0.83695   & 0.08754 & p<0.0000 \\
               & 150\euro/month  & -0.33681  & 0.09101  & p<0.0000\\
               & 200\euro/month  & -0.50014 & (Omitted level) \\
     \bottomrule
  \end{tabular}
  \label{tab6}
\end{table}

\bibliographystyle{unsrt}  


\end{document}